\documentclass[3p,times,twocolumn]{elsarticle}

\usepackage{ecrc}

\volume{00}

\firstpage{1}

\journalname{Nuclear Physics B Proceedings Supplement}

\runauth{M.~K.~Daniel}

\jid{nuphbp}

\jnltitlelogo{Nuclear Physics B Proceedings Supplement}

\usepackage{amssymb}
\usepackage[figuresright]{rotating}

\begin{document}

\begin{frontmatter}



\dochead{}

\title{Lorentz invariance violation with gamma rays}


\author[me]{Michael Daniel}
\author[CTA]{for the CTA Consortium}

\address[me]{Department of Physics, University of Liverpool, Liverpool, L69 7ZE. UK.}
\address[CTA]{https://www.cta-observatory.org/}

\begin{abstract}
The assumption of Lorentz invariance is one of the founding principles 
of Modern Physics and violation of it would have profound implications 
to our understanding of the universe. For instance, certain theories 
attempting a unified theory of quantum gravity predict there could be an 
effective refractive index of the vacuum; the introduction of an energy 
dependent dispersion to photons could in turn lead to an observable 
Lorentz invariance violation signature. Whilst a very small effect on 
local scales the effect will be cumulative, and so for very high energy 
particles that travel very large distances the difference in arrival 
times could become sufficiently large to be detectable. This proceedings will 
look at testing for such Lorentz invariance violation (LIV) signatures 
in the astronomical lightcurves of $\gamma$-ray emitting objects, with 
particular notice being given to the prospects for LIV testing with, the next generation observatory, the Cherenkov Telescope Array.
\end{abstract}


\end{frontmatter}

% \linenumbers

\section{Introduction: Lorentz Invariance and Quantum Gravity}
\label{sec:intro}

Lorentz invariance (LI) is one of the founding principles of the Special Relativity theory of Modern Physics. However it has long been understood that attempting to unify General Relativity (GR) with that other success of Modern Physics, Quantum Mechanics (QM), can in turn lead to deviations from Lorentz symmetry when describing spacetime structure in terms of finite quanta rather an as a continuous lightcone in Minkowski spacetime (\cite{amc2013, ell2013} and references therein). Whilst theories may rely on symmetries, it would not be the first time they would be broken, well known examples being C and CP violation for instance. If LI is only an approximate symmetry of local spacetime and is modified at some scale outside our realm of experience, the Planck scale ($E_\mathrm{QG} \approx E_\mathrm{Pl} \simeq 10^{19}$\,GeV) being a natural one to hypothesise, then there are many models that lead to a vacuum velocity of light that is energy dependent. 

Whilst Quantum Gravity (QG) models are indeed numerous, because the scale of LIV is likely to be so far beyond anything that is feasibly accessible any consequent effect on the observable world would be so correspondingly small\footnote{The most energetic photons recorded are from astrophysical sources and have energies of $\sim$ tens of TeV; for $E_{\gamma} \sim 1$ TeV the correction to the speed of light due to Planck scale linear quantum gravity would be of order $10^{-15}$c.} it can be treated perturbatively and approximated by a dispersion measure that is a simple Taylor expansion, 
\begin{equation}
\label{eq:perturbation}
c^2 p^2 = E_{\gamma}^2\sum_{\alpha}\pm\xi_{\alpha}(E_{\gamma}^{\alpha}/E_{\rm{QG}}^{\alpha})
\end{equation}
where $c$ is the speed of light, $p$ the momentum, $E_{\gamma}$ the energy and $\xi_{\alpha}$ is the correction factor, with the leading linear ($\alpha=1$) and quadratic ($\alpha=2$) terms being those of the most interest. In the linear case, it has been shown that CPT can be violated in effective field theory \cite{ott2013}; however, if CPT is preserved and LI violated it is the quadratic term that would dominate. A positive correction term represents a subluminal change and the negative a superluminal one.

The infinitesimal magnitude of the signature at accessible energy ranges means that these searches require extremely sensitive measurements. Usefully, the minuscule corrections are cumulative and so when photons travel astronomical distances a measurable dispersion in a light curve could be found, although the magnitude of the time delays expected are still only $\delta t \leq 10$\,s/TeV/Gpc for linear Planck scale QG.

\section{Cosmological probes with photons}
\label{sec:probes}
For measuring dispersion due to LIV there are three criteria that an ideal probe should meet:
\begin{itemize}
 \item emit very high energy photons, 
 \item be very distant, 
 \item have very rapid variability.
\end{itemize}
Unfortunately some of these are mutually exclusive, for example very high energy photons will be attenuated by $\gamma + \gamma \rightarrow \mathrm{e}^{+} + \mathrm{e}^{-}$ pair production on the diffuse extragalactic background light, thereby limiting the distance to which these sources will have a detectable signal.

There are probably as many test metrics for time dispersion ($\tau$) as there are QG models (e.g.\cite{bar2012, sca2008, mar2009, vas2013}), but again they reduce down to a simple search of arrival-time correction, $\delta t_i$, on a photon $i$ of energy $E_i$ such that 
%\begin{equation}
%\label{eq:cancel}
$ \delta t_i = -\tau E_i^\alpha $.
%\end{equation}
The {\em dispersion cancellation} algorithm cycles through a range of possible $\tau$, looking for the value that extremises the metric. If $\tau \neq 0$ dispersion must be present. An advantage of this approach is that it makes no {\em a priori} assumptions on the nature of the lightcurve apart from the inevitable hypothesis of simultaneity of emission of photons of all energies at the source. Since these are sites of particle acceleration that is not necessarily an accurate assumption. For this reason it is best to search for a LIV signature in a number of sources so that they are subject to different intrinsic physical processes and at varying redshifts. A LIV signature should scale with distance, whereas it is unlikely (but not impossible \cite{rei2010}) the same would be true of the intrinsic processes.
%(see examples in \cite{ellis08}, \cite{magicqg} and \cite{scargle08})

% - time of flight
%   -- GRB Fermi & Julien
%   -- AGN ...
% - Pulsars
%   -- veritas/otte
% - Gamma ray horizon
%   -- Fairbairn et al

\subsection{Gamma ray bursts (GRBs)}
%First suggested by \cite{Amelino-Camelia98}
Gamma ray bursts show the fastest variability\footnote{a short duration GRB has a duration of $<2$\,s} and have the furthest distance of known $\gamma$-ray sources, but not the highest energy. Observed by the Fermi-LAT satellite~\cite{FermiSpecs}, GRB\,090510 has provided for some of the most constraining limits yet, with a limit above the Planck scale for linear scale LIV induced dispersion \cite{FermiLimits, vas2013}. If detected by a ground based instrument the photon statistics at the highest energies would increase significantly making for even more constraining limits, but the small field of view of most of these instruments make this challenging to catch them serendipitously: GRBs also suffer in that they are unpredictable in location and distance, so it is difficult to build up statistics with them as individual sources.

\subsection{Active Galactic Nuclei (AGN)}
The jets from AGN make for repeatable and even more energetic sources of photons than GRBs, but at the expense of longer ($\sim$ minute) scale variability features in the light curve. Current LIV limits are just below the Planck scale on the linear term with current generation instruments observations of AGN flares \cite{xcol2155, likelihood2155, magicqg}, but the next generation facility the Cherenkov Telescope Array (CTA) \cite{CTA} will have the sensitivity to match or beat current GRB limits. Even if the AGN lightcurve does not have sufficiently rapid features to determine dispersion for any single flaring episode, a LIV induced dispersion will mean that higher energy photons will always arrive shifted with respect to lower energy ones in the lightcurve. The accumulation of long term monitoring data means that we can still potentially determine time delays at high confidence, e.g. through the use of cross-power spectral analysis methods \cite{dor2013}. The underlying rapid varying features monitored over long periods serving to further increase the chance of detection. This will be the first time that routine AGN observations, i.e. not on exceptional flux levels, will provide us with such LIV constraints.

\subsection{Pulsars}
When it comes to testing the quadratic term for LIV, having a very high energy component compensates for a lack of distance. Pulsars represent a fast varying, relatively well understood source population with very different intrinsic source physics processes to GRBs and AGN. The Crab pulsar has a pulsed VHE component to its spectrum up to hundreds of GeV and little evidence of a cut-off (within event statistics). Whilst the LIV limits from current generation instruments are presently inferior to those from AGN and GRBs \cite{zit2013}, a millisecond pulsar observed at 1\,TeV with CTA has the potential to place stringent limits (even above the Planck scale on the linear term). As a pulsar has a very well measured lightcurve profile it also makes for an interesting source to test for any lightcurve broadening that might occur from a polarisation dependent superluminal correction (see e.g.~\cite{superluminal, IntegralLimits}).

\subsection{The gamma ray horizon}
If LIV modifies the dispersion relation for $\gamma$ rays it could also affect the kinematics in the pair production process that attenuates the VHE signal as it travels through the diffuse extragalactic background light, changing the cross-section and allowing VHE photons (up to hundreds of TeV) to be detected that would not normally be expected in deep observations of suitably hard spectrum distant AGN/GRBs \cite{kif1999, fai2014}.

\section{Summary}
\label{sec:summary}
Testing for Lorentz invariance violation (LIV) with TeV and higher energy particles can either verify GR at an entirely new sensitivity level, or inform the development of new quantum gravity (QG) models. 
The negligible LIV effects at accessible energies would become noticeable only when accumulated after travelling astronomical distances, which means that it is distant, rapidly variable sources like GRBs and AGN that provide the most sensitive probes to LIV effects. 
One of the main caveats in studying LIV with current generation $\gamma$-ray instruments is the difficulty in disentangling intrinsic source physics dispersion from propagation induced effects. The sensitivity of CTA will enable us to overcome this through the observation of many sources as a function of redshift with high statistics and to measure for effects of both modified dispersion and modified fundamental interactions giving new insight into both source and fundamental physics processes. This will enable, in a completely model independent way, to test the functional form between the delays as a function of distance, i.e. linear, quadratic or other.

Extraordinary claims require extraordinary evidence and so multi-object, multi-wavelength and multi-messenger observations are needed to gather evidence in the case for or against LIV. As pointed out in \cite{ell2013} individual particle's characteristics (such as charge) can determine the kind of interactions that determine the amount of dispersion, if any, it can experience on its journey through space-time and so be a crucial factor in discriminating between scenarios. 

We gratefully acknowledge support from the agencies and organizations 
listed under Funding Agencies at this website: http://www.cta-observatory.org/



%% References

\nocite{*}
\bibliographystyle{elsarticle-num}
\bibliography{mkd-liv}



%\bibitem{amc2013} Amelino-Camelia~G. LRR 16, (2013) 5
%\bibitem{ell2013} Ellis~J. \& Mavromatos~N.~E. APh 43 (2013) 50
%\bibitem{rei2010} Reimer~A. Nu.Ph.B 203 (2010) 33
%\bibitem{ott2013} Otte~N. et al. arXiv:1305.0264
%\bibitem{dor2013} Doro~M. et al. APh 43 (2013) 189
%\bibitem{fai2014} Fairbairn~M. et al. JCAP 06 (2014) 005
%\bibitem{kif1999} Kifune~T. ApJ 518 (1999) 21



% \end{thebibliography}

\end{document}